\documentclass[11pt]{article}
\usepackage{hhline}
\usepackage{clrscode}

\usepackage{dustinMacro}
\usepackage{url}
\usepackage{empheq}

\usepackage{color} 
\usepackage[latin1]{inputenc}
\usepackage{nicefrac} 
\usepackage{paralist} 

\usepackage{enumitem}
\setlist{nolistsep}

\usepackage{boxedminipage} 
\usepackage{fullpage} 
\usepackage[pdfstartview=FitH,colorlinks,linkcolor=blue,filecolor=blue,citecolor=blue,urlcolor=blue,pagebackref=true]{hyperref}
\usepackage{upgreek} 
\usepackage{tikz}
\usepackage{times}
\usetikzlibrary{arrows,shapes}

\newcommand{\commentout}[1]{}

\topmargin by -\headsep \textheight 9.1in \oddsidemargin 0pt
\evensidemargin \oddsidemargin \marginparwidth 0.4 in \textwidth
6.6in

\newcommand{\negl}{\mathrm{negl}}

\newcommand{\nmcom}{\ensuremath{\langle C, R \rangle}\xspace}

\newcommand{\SIM}{{\sf SIM}}

\def \arrowlength #1 {\gdef \@arrowlength{#1}}
\newcommand{\rsend}[1]{{\ensuremath{\underrightarrow{\text{\makebox[\@arrowlength][c]{\ensuremath{#1}}}}}}}
\newcommand{\lsend}[1]{{\ensuremath{\underleftarrow{\text{\makebox[\@arrowlength][c]{\ensuremath{#1}}}}}}}

\newcommand{\Tr}{\mathrm{Tr}}

\newcommand{\tpath}{\mathrm{path}}

\newcommand{\rt}{\lambda}
\newcommand{\zo}{{\{0,1\}}}
\newcommand{\qmax}{q_{\mathrm{max}}}
\newcommand{\abortqueue}{\proc{AbortQueue}}
\newcommand{\abortleaf}{\proc{AbortLeaf}}

\providecommand{\ie}{\emph{i.e.,} }
\providecommand{\eg}{\emph{e.g.,} }

\providecommand{\myparab}[1]{\smallskip\noindent\textbf{#1} }

\newtheorem{assumption}{Assumption}            %
\newtheorem{proposition}{Proposition}

\newcommand{\BA}{\begin{assumption}}   \newcommand{\EA}{\end{assumption}}

\arrowlength{2.5in}
\def \ShowAuthNotes{1}
\ifnum\ShowAuthNotes=1
\newcommand{\authnote}[2]{{ \textbf{ [#1's Note:} {\em #2} \textbf{]} }}
\else
\newcommand{\authnote}[2]{}
\fi

\newcommand{\Knote}[1]{{\color{red}\authnote{Kai-Min}{#1}}}

\newcommand{\Znote}[1]{{\color{red}\authnote{Zhenming}{#1}}}

\begin{document}
\thispagestyle{empty}
\begin{titlepage}

  \title{Statistically-secure ORAM with $\tilde{O}(\log^2 n)$ Overhead}

 \author{Kai-Min Chung\thanks{Cornell University.
     \texttt{\{chung,rafael\}@cs.cornell.edu} \newline
Chung is supported in part by NSF Award CCF-1214844 and Pass' Sloan Fellowship. \newline
Pass is supported in part by a Alfred P. Sloan Fellowship,
  Microsoft New Faculty Fellowship, NSF Award CNS-1217821, NSF CAREER
 Award CCF-0746990, NSF Award CCF-1214844, AFOSR YIP Award
  FA9550-10-1-0093, and DARPA and AFRL under contract FA8750-11-2-
  0211. The views and conclusions contained in this document are those
  of the authors and should not be interpreted as representing the
  official policies, either expressed or implied, of the Defense
  Advanced Research Projects Agency or the US Government.} ~~~ Zhenming Liu\thanks{Princeton University. \texttt{zhenming@cs.princeton.edu}}      ~~~ Rafael Pass\footnotemark[1]    \vspace{4ex}}

\maketitle
\begin{abstract}

We demonstrate a simple, statistically secure, ORAM with computational
overhead $\tilde{O}(\log^2 n)$; previous ORAM protocols achieve only computational security (under computational assumptions) or require $\tilde{\Omega}(\log^3 n)$ overheard. An additional benefit of our ORAM is its conceptual simplicity, which makes it easy to implement in both software and (commercially available) hardware. 

Our construction is based on recent ORAM constructions due to Shi, Chan, 
Stefanov, and Li~(Asiacrypt 2011) and Stefanov and Shi~(ArXiv 2012),
but with some crucial modifications in the algorithm that simplifies
the ORAM and enable our analysis. A central component in our analysis
is reducing the analysis of our algorithm to a ``supermarket''
problem; of independent interest (and of importance to our analysis,)
we provide an upper bound on the rate of ``upset'' customers in the
``supermarket'' problem.

\end{abstract}

\thispagestyle{empty}
\end{titlepage}
\maketitle

\newcommand{\prob}{{\rm Pr}}
\newcommand{\SIMnml}[1]{\ensuremath{\SIM_{\nmcom}^{A}{(#1)}}}
\newcommand{\WI}{{\cal WI}} 
\newcommand{\ZK}{{\cal ZK}}
\newcommand{\srWI}{{\cal \mathsf{sr}WI}}
\newcommand{\rWI}{{\cal \mathsf{r}WI}}

\section{Introduction}
In this paper  we consider constructions of \emph{Oblivious RAM (ORAM)}
\cite{G87, GO96}. Roughly speaking, an ORAM enables executing a
RAM program while hiding the access pattern to the memory. 
ORAM have several fundamental applications (see e.g. \cite{GO96, OS97} for
further discussion). Since the seminal works for Goldreich \cite{G87} and
Goldreich and Ostrovksy \cite{GO96}, constructions of ORAM have been
extensively studied (see e.g., \cite{WS08,WSC08,A10,PR10,GM11,DMN11,SCSL11,BMP11,GMOT12,SSS12,KLO12}.) While the original constructions only enjoyed
``computational security'' (under the the assumption
that one-way functions exists) and required a computational overhead
of $\tilde{O}(\log^3 n)$, more recent works have overcome both of
these barriers, but only individually. State of the art ORAMs satisfy
either of the following:
\begin{itemize}
\item  An overhead of $\tilde{O}(\log^2 n)$\footnote{The
    best protocol achieves $O(\log^2 n/\log \log n)$.}, but only
    satisfies computational security, assuming the existence of
    one-way functions. \cite{PR10,GM11,KLO12}
\item  Statistical security, but have an 
    overhead of $O(\log^3 n)$. \cite{A10,DMN11,SCSL11,GGHJRW13,CP13ORAM}.
\end{itemize}
A natural question is whether both of these barriers can be
simultaneously overcome; namely, does there exists a statistically
secure ORAM with only $\tilde{O}(\log^2 n)$ overhead? In this work we
answer this question in the affirmative, demonstrating the existence
of such an ORAM.

\begin{Thm}\label{thm:main}
There exists a statistically-secure ORAM with $\tilde O(\log^2(n))$ worst-case
computational overhead, constant memory overhead, and CPU cache size $\poly\log(n)$,  where 
$n$ is the memory size. 
\end{Thm}

An additional benefit of our ORAM is its conceptual simplicity, which
makes it easy to implement in both software and (commercially
available) hardware. (A software implementation is available from the
authors upon request.)

\paragraph{Our ORAM Construction}
A conceptual breakthrough in the construction of ORAMs appeared in the
recent work of Shi, Chan, 
Stefanov, and Li \cite{SCSL11}. This work demonstrated a statistically secure
ORAM with overhead $O(\log^3 n)$ using a new ``tree-based''
construction framework, which admits significantly simpler (and
thus easier to implemented) ORAM constructions (see also \cite{GGHJRW13,CP13ORAM} for
instantiations of this framework which additionally enjoys an extremely
simple proof of security).

On a high-level, each memory cell $r$ accessed by the original RAM
will be associated with a random leaf $pos$ in a binary tree; the
position is specified by a so-called ``position map'' $Pos$.
Each node in the tree consists of a ``bucket'' which stores up to
$\ell$ elements. The content of memory cell $r$ will be found inside
one of the buckets along the path from the root to the leaf $pos$;
originally, it is put into the root, and later on, the content gets
``pushed-down'' through an eviction procedure---for instance, in the ORAM of
\cite{CP13ORAM} (upon which we rely), the eviction procedure consists of
``flushing'' down memory contents along a random path, while ensuring
that each memory cell is still found on its appropriate path from the
root to its assigned leaf. (Furthermore, each time the content
of a memory cell is accessed, the content is removed from the tree,
the memory cell is assigned to a new random leaf, and the content
is put back into the root).

In the work of \cite{SCSL11} and its follow-ups \cite{GGHJRW13,CP13ORAM}, for the
analysis to go through, the bucket size $\ell$ is required to be
$\omega(\log n)$. Stefanov and Shi~\cite{SS12} recently provided a different
instantiation of this framework which only uses \emph{constant size}
buckets, but instead relies on a \emph{single} $\poly\log n$ size
``stash'' into which potential ``overflows'' (of the buckets in the
tree) are put; Stefanov and Shi conjectured (but did not prove) security of such a
construction (when appropriately evicting elements from the ``stash''
along the path traversed to access some memory cell).\footnote{Although
  different, the
  ``flush'' mechanism in \cite{CP13ORAM} is inspired by this eviction method.}

In this work, we follow the above-mentioned approaches, but with the
following high-level modifications:
\begin{itemize}
\item We consider a binary tree where the bucket size of all
  \emph{internal} buckets is $O(\log \log n)$, but all the leaf nodes
  still have bucket size $\omega(\log n)$.
\item As in \cite{SS12}, we use a ``stash'' to store potential
  ``overflows'' from the bucket. In our ORAM we refer to this as a
  ``queue'' as the main operation we require from it is to insert and
  ``pop'' elements (as we explain shortly, we additionally need to be able
  to find and remove any particular element from the queue; this can
  be easily achieved using a standard hash table). Additionally,
  instead of inserting memory cells directly into the tree, we insert
  them into the queue. When searching for a memory cell, we first
  check whether the memory cell is found in the queue (in which case
  it gets removed), and if not, we search for the memory cell in the binary tree
  along the path from the root to the position dictated by the position map. 
\item Rather than just ``flushing'' once (as in \cite{CP13ORAM}), we
  repeat the following procedure ``pop and random flush'' procedure twice. 
\begin{itemize}
\item We ``pop'' an element from the queue into the
  root.
\item Next, we flush according to a \emph{geometrically distributed} random variable with
  expectation 2.\footnote{Looking forward, our actual flush is a
    little bit different than the one in \cite{CP13ORAM} in that we only
    pull down a \emph{single} element between any two consecutive nodes
    along the path, whereas in \cite{CP13ORAM} \emph{all} elements that can be
    pulled down get flushed down.} 
\end{itemize}
\end{itemize}
We demonstrate that such an ORAM construction is both (statistically)
secure, and only has $\tilde{\Omega}(\log^2 n)$ overhead.

\paragraph{Our Analysis}
The key element in our analysis is reducing the security of our ORAM
to a ``supermarket'' problem. Supermarket problems were introduced by
Mitzenmacher \cite{M01} and have seen been well-studied (see e.g.,
\cite{M01,VDK96,MV99,DB02,MPS02}). We here consider a simple version of a supermarket problem,
but ask a new question: what is the rate of ``upset'' customers in a
supermarket problem:
There are $D$ cashiers in the supermarket, all
of which have empty queues in the beginning of the day. At each time
step $t$: with probability $\alpha < 1/2$ a new customer arrives and
chooses a random cashier\footnote{Typically, in supermarket problems
  the customer chooses $d$ random cashiers and picks the one with the
  smallest queue; we here focus on the simple case when $d=1$.}
(and puts himself in that cashiers
queue); otherwise (i.e., with probability $1-\alpha$) a random cashier is chosen
that ``serves'' the first customer in its queue (and the queue size is
reduced by one). We say that a customer is \emph{upset} is he
chooses a queue whose size exceeds some bound $\varphi$.
What is the rate of upset customers?\footnote{Although we here
  consider a discrete-time version of the supermarket problem (since
  this is the most relevant for our application), as we remark in
  Remark \ref{rmk:discrete-to-cont}, our results apply also to the more commonly studied continuous-time setting.}

We provide an upper bound on the rate of upset customers relying on
Chernoff bounds for Markov chains \cite{Gillman93,Kahale97,Lezaud04,CLLM12}---more specifically, we
develop a variant of traditional Chernoff bounds for Markov chains
which apply also with ``resets'' (where at each step, with some small
probability, the distribution is reset to the stationary distribution
of the Markov chain), which may be of independent interest, and show
how such a Chernoff bound can be used in a rather straight-forward way
to provide a bound on the number of upset customers.

Intuitively, to reduce the security of our ORAM
to the above-mentioned supermarket problem, each cashier corresponds to a
bucket on some particular level $k$ in the tree, and the bound
$\varphi$ corresponds to the bucket size, customers correspond to
elements being placed in the buckets, and upset customers overflows.
Note that for this translation to work it is important that the number
of flushes in our ORAM is geometrically distributed---this ensures that we can
view the sequence of opertaions (i.e., ``flushes'' that decrease
bucket sizes, and ``pops'' that increase bucket sizes) as
independently distributed as in the supermarket problem. 

\paragraph{Independent Work}
In a very recent independent work, Stefanov, van Dijk, Shi, Fletcher, Ren, Yu, and Devadas~\cite{SvDSFRYD13} prove
security of the conjectured Path ORAM of \cite{SS12}. 
This yields a ORAM with overhead 
$O(\log^2  n)$, whereas our ORAM has overhead 
$O(\log^2 n \log \log n)$).
On the other hand, the data structure required to implement our queue
is simpler than the one needed to implement the ``stash'' in the Path
ORAM construction. More precisely, we simply need a standard queue and
a standard hash table (both of which can be implemented using
commodity hardware), whereas the ``stash'' in \cite{SS12,SvDSFRYD13}
requires using a data structure that additionally supports ``range queries'', and
thus a binary search tree is needed, which may make implementations
more costly.
We leave a more complete exploration of the benefits of the
different approaches for future work. 

\section{Preliminaries}
A Random Access Machine (RAM) with memory size $n$ consists of a CPU
with a small size cache (e.g., can store  a constant or $\poly\log(n)$ number of words) and an ``external'' memory of size 
$n$. To simplify notation, a word is either $\bot$ or a $\log n$ bit string. 

The CPU executes a program $\Pi$ (given $n$ and some input $x$) that can access the memory by a
$Read(r)$ and $Write(r,val)$ operations where $r \in [n]$ is an index to a memory
location, and $val$ is a word (of size $\log n$). 
The sequence of memory cell
accesses by such read and write operations is referred to as the \emph{memory access pattern} of $\Pi(n,x)$
and is denoted $\tilde{\Pi}(n,x)$.
(The CPU may also
execute ``standard'' operations on the registers, any may generate outputs).

Let us turn to defining an {\em Oblivous RAM Compiler}. This notion
was first defined by Goldreich \cite{G87} and Goldreich and Ostrovksy
\cite{GO96}.
We recall a more succinct variant of their
definition due to~\cite{CP13ORAM}.
\begin{Definition}
A polynomial-time algorithm $C$ is an \emph{Oblivious RAM (ORAM) compiler} with computational overhead $c(\cdot)$ and
memory overhead $m(\cdot)$, 
if $C$ given $n\in N$ and a deterministic RAM
program $\Pi$ with memory-size $n$ outputs a program $\Pi'$ 
with memory-size $m(n)\cdot n$ such that for any input $x$, the running-time of $\Pi'(n,x)$ is
bounded by $c(n)\cdot T$ where $T$ is the running-time of $\Pi(n,x)$, and
there exists a negligible function
$\mu$ such that 
the following properties hold:
\begin{itemize}

\item {\bf Correctness:} 
For any $n\in N$ and any string $x \in \{0,1\}^*$, with probability at least 
$1-\mu(n)$, $\Pi(n,x) = \Pi'(n,x)$.
\item {\bf Obliviousness:}
For any
two programs $\Pi_1$, $\Pi_2$, any $n\in N$ and any two
inputs $x_1,x_2 \in \bitset^*$ if $|\tilde{\Pi}_1(n,x_1)| =
|\tilde{\Pi}_2(n,x_2)|$, then $\tilde{\Pi}'_1(n,x_1)$ is
$\mu$-close to $\tilde{\Pi}'_2(n,x_2)$ in statistical distance, where
$\Pi'_1 = C(n,\Pi_1)$ and $\Pi'_2 = C(n,\Pi_2)$.
\end{itemize}
\end{Definition}
Note that the above definition (just as the definition of \cite{GO96})
only requires an oblivious compilation of \emph{deterministic}
programs $\Pi$. This is without loss of generality: we can always view
a randomized program as a deterministic one that receives random coins
as part of its input.

\section{Algorithm for the ORAM.}

Our ORAM data structure serves as a ``big'' memory table of size $n$ and exposes the following two interfaces. 

\begin{itemize}
\item $\proc{Read}(r)$: the algorithm returns the value of memory cell $r \in [n]$. 
\item $\proc{Write}(r, v)$: the algorithm writes value $v$ to memory cell $r$. 
\end{itemize}

We start assuming that the ORAM is executed on a CPU with cache size is
$2n/\alpha + o(n)$ (in words) for a suitably large constant $\alpha$
(the reader may imagine $\alpha = 16$). Following the framework in
\cite{SCSL11}, we can then reduce the cache
size to $O(\mathrm{poly}\log n)$ by recursively applying the
ORAM construction; we provide further details on this transformation
at the end of the section.
 
In what follows, we group each consecutive $\alpha$ memory cells
in the RAM into a \emph{block} and will thus 
have $n/\alpha$ blocks in total. We also index the blocks in the natural way, \ie the block that contains the first $\alpha$ memory cells in the table has index $0$ and in general the $i$-th block contains memory cells with addresses from $\alpha i$ to $\alpha (i + 1) - 1$.

Our algorithm will always be operating at the block level, \ie memory cells in the same block will always be read/written together. 
In addition to the content of its $\alpha$ memory cells, each block is associated with two extra pieces of information. First, it stores the index $i$ of the block. 
Second, it stores a ``position'' $p$ that specify it's storage
``destination'' in the external memory, which we elaborate upon in the forthcoming paragraphs. In other words, a block is of the form $(i,p,val)$, where $val$ is the content of its $\alpha$ memory cells.

Our ORAM construction relies on the following three main components. 

\begin{enumerate}
\itemsep0em 
\item \textbf{A full binary tree at the in the external memory} that serves as the primary media to store the data. 
\item \textbf{A position map in the internal cache} that helps us to
  search for items in the binary tree. 
\item \textbf{A queue in the internal cache} that is the secondary venue to store the data. 
\end{enumerate}

We now walk through each of the building blocks in details. 

\myparab{The full binary tree $\Tr$.} The depth of this full binary
tree is set to be the smallest $d$ so that the number of leaves $L =
2^d$ is at least  $2(n/\alpha)/ (\log n \log \log n)$ (\ie, $L/2 <
2(n/\alpha)/ (\log n \log \log n) \leq L$).(In
  \cite{SCSL11,CP13ORAM} the number of leaves was set to $n/\alpha$;
  here, we instead follow \cite{GGHJRW13} and make the tree slightly
  smaller---this makes the memory overhead smaller.)
We index nodes in the tree by a binary
string of length at most $d$, where the root is indexed by the empty
string $\lambda$, and each node indexed by $\gamma$ has left and right
children indexed $\gamma 0$ and $\gamma 1$, respectively.
Each node is associated with a \emph{bucket}. 
A bucket in an internal node can store up to $\ell$ blocks, and a bucket in a leaf can store up to $\ell'$ blocks, where $\ell$ and $\ell'$ are parameters to be determined later.
The tree shall support the following two atomic operations:
\begin{itemize}
\item $\proc{Read}(\mbox{Node: }v)$: the  tree will return all the blocks in the bucket associated with $v$ to the cache. 
\item $\proc{Write}(\mbox{Node: }v, \mbox{Blocks: } \vec{b} )$: the
  input is a node $v$ and an array of blocks $\vec{b}$ (that will fit
  into the bucket in node $v$). This operation will replace the bucket in the node $v$ by $\bar{b}$.   
\end{itemize}

\myparab{The position map $P$.} This data structure is an array that maps the indices of the blocks to leaves in the full binary tree.
Specifically, it supports the following atomic operations:
\begin{itemize}
\item $\proc{Read}(i)$: this function returns the position $P[i] \in [L]$ that corresponds to the block with index $i \in [n/\alpha]$. 
\item $\proc{Write}(i, p)$: this function writes the position $p$ to $P[i]$. 
\end{itemize}

\myparab{The queue $Q$.}  This data structure stores a queue of blocks with maximum size $\qmax$,  a parameter to be determined later, and supports the following three atomic operations:
\begin{itemize}
\item $\proc{Insert}(\mathrm{Block }\  b)$: insert a block $b$ into the queue. 
\item $\proc{PopFront}()$: the first block in the queue is popped and returned. 
\item $\proc{Find}(\mbox{int: } i, \mbox{word: } p)$: if there is a block $b$ with index $i$ and position $p$ stored in the queue, then $\proc{Find}$ returns $b$ and deletes it from the queue; otherwise, it returns $\bot$. 
\end{itemize}

Note that in addition to the usual $\proc{Insert}$ and $\proc{PopFront}$ operations, we also require the queue to support a $\proc{Find}$ operation that finds a given block, returns and deletes it from the queue. This operation can be supported using a standard hash table in conjunction with the queue. We mention that all three operations can be implemented in time less than $O(\log n \log \log n)$, and discuss the implementation details in Appendix~\ref{sec:impl}.

\myparab{Our Construction.} We now are ready to describe our ORAM construction, which relies the above atomic operations. Here, we shall
focus on the read operation. The algorithm for the write operation is analogous. 

For two nodes $u$ and $v$ in $\Tr$, we use $\tpath(u, v)$ to denote  the (unique) path connecting $u$ and $v$.  
Throughout the life cycle of our algorithm we maintain the following \emph{block-path} invariance.
\begin{quote}
{\bf Block-path Invariance}: \emph{ For any index $i \in [n/\alpha]$,
  there exists at most a single block $b$ with index $i$ that is
  located either in $\Tr$ or in the queue. %
When it is in the tree, it will be in the bucket of one of the nodes on $\tpath(\rt,P[i])$. Additionally, $b$ has position $p = P[i]$.}
\end{quote}  

We proceed to describe our $\proc{Read}(r)$ algorithm. At a high-level, $\proc{Read}(r)$ consists of two sub-routines $\proc{Fetch}()$ and $\proc{Dequeue}()$, where we executes $\proc{Fetch}()$ once, and then executes $\proc{Dequeue}()$ \emph{twice}. Roughly, $\proc{Fetch}()$ fetches the block $b$ that contains the memory cell $r$ from either $\tpath(\rt,P[\lfloor r/\alpha  \rfloor])$ in $\Tr$ or in $Q$,  then returns the value of memory cell $r$, and finally inserts the block $b$ to the queue $Q$. On the other hand, $\proc{Dequeue}()$ pops one block $b$ from $Q$, inserts $b$ to the root $\rt$ of $\Tr$ (provided there is a room), and performs a \emph{random} number of ``\proc{Flush}'' actions that gradually moves blocks in $\Tr$ down to the leaves. 

\begin{description}
\item [Fetch:] Let $i = \lfloor r/\alpha \rfloor$ be the index of the block $b$ that contains the $r$-th memory cell, and $p = P[i]$ be the current position of $b$. If $P[i] = \bot$ (which means that the block is not initialized yet), let $P[i]  \leftarrow [L]$ be a uniformly random leaf, create a block $b = (i, P[i], \vec{0})$, and insert $b$ to the queue $Q$. Otherwise, $\proc{Fetch}$ performs the following actions in order.

\myparab{Fetch from tree $\Tr$ and queue $Q$:} Search the block $b$ with index $i$ along $\tpath(\rt, p)$ in $\Tr$ by reading all buckets in $\tpath(\rt, p)$ once and writing them back. If such a block is found, save it and write back a dummy block; otherwise, search the block $b$ with index $i$ and position $p$ in the queue $Q$ by invoking $\proc{Find}(i,p)$. By the block-path invariance, we must find the block $b$.

\myparab{Update position map $P$.}Let $P[i] \leftarrow [L]$ be a uniformly random leaf, and update the position $p = P[i]$ in $b$.

\myparab{Insert to queue $Q$:}Insert the block $b$ to $Q$.

\item [Dequeue:] This sub-routine consists of two actions $\proc{Put-Back}()$ and $\proc{Flush}()$. It starts by executing $\proc{Put-Back}()$ once, and then performs a \emph{random} number of $\proc{Flush}()$es as follows: Let $C\in \zo$ be a biased coin with $\Pr{ C = 1 } = 2/3$. It samples $C$, and if the outcome is $1$, then it continues to perform one $\proc{Flush}()$ and sample another independent copy of $C$, until the outcome is $0$. (In other words, the number of $\proc{Flush}()$ is a geometric random variable with parameter $2/3$.)

\myparab{Put-Back:} This action moves a block from the queue, if any, to the root of $\Tr$.
Specifically, we first invoke a $\proc{PopFront}()$. If
$\proc{PopFront}()$ returns a block $b$ 
then add it $b$ to $\rt$ .


\myparab{Flush}: This procedure selects a random path (namely, the path connecting the root to a random leaf $p^* \leftarrow \zo^{d}$) on the tree and tries to move the blocks along the path 
down subject to the condition that the block always finds themselves
on the appropriate path from the root to their assigned leaf node
(see the block-path invariance condition). Let $p_0 (= \rt) p_1...p_{d  }$  be the nodes along $\tpath(\rt,p^*)$. We traverse the path  while carrying out the following operations 
for each node $p_i$ we visit: in node $p_i$, find the block that can
be ``pulled-down'' as far as possible along the path $\tpath(\rt,p^*)$
(subject to the block-path invariance condition), and pull it down to $p_{i+1}$.
For $i<d$, if there exists some $\eta \in \{0,1\}$ such that $p_i$
contains more than $\ell/2$ blocks that are assigned to leafs of the
form $p_i||\eta||\cdot$, then select an arbitrary such block $b$,
remove it from the bucket $p_i$ and invokes an $\proc{Overflow}(b)$
procedure, which re-samples a uniformly random
position for the overflowed block $b$ and inserts it back to the queue
$Q$. (See Figure~\ref{fig:flush} and \ref{fig:overflow} in Appendix for the pseudocode)

\end{description}

Finally, the algorithm aborts and terminates if one of the following two events happen throughout the execution.
\begin{description}
\item [Abort-queue]: If the size of the queue $Q$ reaches $\qmax$, then the algorithm aborts and outputs $\abortqueue$.
\item [Abort-leaf]: If the size of any leaf bucket reaches $\ell'$ (i.e., it becomes full), then the algorithm aborts and outputs $\abortleaf$.
\end{description}

This completes the description of our $\proc{Read}(r)$  algorithm; the
$\proc{Write}(r,v)$ algorithm is defined in identically the same way,
except that instead of inserting $b$ into the queue $Q$ (in the last
step of $\proc{Fetch}$), we insert a modified $b'$ where the content
of the memory cell $r$ (inside $b$) has been updated to $v$.

It follows by inspection that the block-path invariance is preserved by our construction. Also, note that in the above algorithm, $\proc{Fetch}$ increases the size of the queue $Q$ by $1$ and $\proc{Put-back}$ is executed twice which decreases the queue size by $2$. On the other hand, the $\proc{Flush}$ action may cause a few $\proc{Overflow}$ events, and when an $\proc{Overflow}$ occurs, one block will be removed from $\Tr$ and inserted to $Q$. Therefore, the size of the queue changes by minus one plus the number of $\proc{Overflow}$ for each $\proc{Read}$ operation. The crux of our analysis is to show that the number of $\proc{Overflow}$ is sufficiently small in any given (short) period of time, except with negligible probability.

We remark that
throughout this algorithm's life cycle, there will be at most $\ell-2$
non-empty blocks in each internal node except when we invoke
$\proc{Flush}(\cdot)$, in which case some intermediate states will
have  $\ell-1$ blocks in a bucket (which causes an invokation of $\proc{Overflow}$).

\myparab{Reducing the cache's size.} 
We now briefly describe how the cache can be reduced to $\poly\log(n)$. We will set the queue size $\qmax = \poly\log(n)$ (specifically, we can set $\qmax = O(\log^{2+\eps} n)$ for an arbitrarily small constant $\eps$).
The key observation here
is that the position map shares the same set of interfaces with our ORAM data structure. Thus, we may 
substitute the position map with a (smaller) ORAM of size $[n/\alpha]$. By recursively substituting 
the position map $O(\log n)$ times,  the size of the cache will reduce to $\mathrm{poly}\log n$.

\myparab{Efficiency and setting parameters.}\label{sec:simpleanalysis} 
By inspection, it is not hard to see that the runtime of our $\proc{Read}$ and $\proc{Write}$ algorithms is $O(\ell\log^2n  + \ell' \log n)$. Also, note that the position map of the base construction has size $O((\ell + \ell') \cdot L) = O((\ell + \ell') \cdot (n/\alpha)/ (\log n \log \log n))$, and each recursive level has a position map that is a constant factor smaller. Thus, the overall external memory required by our ORAM construction remains $O((\ell + \ell') \cdot (n/\alpha)/ (\log n \log \log n))$. To achieve the claims efficiency in Theorem~\ref{thm:main}, we set $\ell = O(\log \log n)$ and $\ell' = O(\log n \log \log n)$.


\section{Security of our ORAM} \label{sec:security}

The following observation is central to the security of  our ORAM construction
(and an appropriate analogue of it was central already to the
constructions of ~\cite{SCSL11,CP13ORAM}):
\begin{quote}
{\em {\bf Key observation:} Let $X$ denote the sum of two independent geometric random variables with mean $2$. Each $Read$ and $Write$ operation traverses the tree along $X+1$ randomly
chosen paths, \emph{independent} of the history of operations so far.}
\end{quote} 
The key observation follows from the facts that (1) just as in the
schemes of ~\cite{SCSL11,CP13ORAM}, each position in the
position map is used exactly once in a traversal (and before this
traversal, no information about the position is used in determining
what nodes to traverse), and (2) we invokes the \proc{Flush} action $X$ times and the flushing, by definition, traverses a
random path, independent of the history.

Armed with the key observation, the security of our construction reduces to show that our ORAM program does not  aborts except with  negligible probability, which follows by the following two lemmas.

\begin{Lem}\label{lem:leaf} Given any program $\Pi$, let $\Pi'(n,x)$ be the compiled program using our ORAM construction. We have
$$\Pr{ \abortleaf } \leq \negl(n).$$
\end{Lem}

\begin{Lem}\label{lem:queue} Given any program $\Pi$, let $\Pi'(n,x)$ be the compiled program using our ORAM construction. We have
$$\Pr{ \abortqueue } \leq \negl(n).$$
\end{Lem}

The proof of Lemma \ref{lem:leaf} is found in the Appendix and follows
by a direct application of the (multiplcative) Chernoff bound.
The proof of Lemma \ref{lem:queue} is significantly more interesting.
Towards proving it, in Section \ref{sec:simpleanalysis} we consider a simple variant of a
``supermarket'' problem (introduced by Mitzenmacher\cite{M01}) and show how to reduce Lemma
\ref{lem:queue} to an (in our eyes) basic and natural question that
seems not to have been investigated before.

\section{Proof of Lemma \ref{lem:queue}}\label{sec:simpleanalysis}
We here prove Lemma~\ref{lem:queue}: in Section~\ref{sec:smintro} we
consider a notion of ``upset'' customers in a supermarket
problem~\cite{M01,VDK96,ELZ86}; in Section~\ref{sec:step1} we show how
Lemma~\ref{lem:queue} reduced to obtaining a bound on the rate of
upset customers, and in Section~\ref{sec:step2} we provide an upper bound
on the rate of upset customers.

\subsection{A Supermarket Problem}\label{sec:smintro}
In a supermarket problem, there are $D$ cashiers in the supermarket, all of which have empty queues in the beginning of the day. At each time step $t$,
\begin{itemize}
\item With probability $\alpha < 1/2$, an \emph{arrival} event happens, where a new customer arrives. The new customer chooses $d$ uniformly random cashiers and join the one with the shortest queue.
\item Otherwise (\ie with the remaining probability $1-\alpha$), a \emph{serving} event happens: a random cashier is chosen that ``serves'' the first customer in his queue and the queue size is reduced by one; if the queue is empty, then nothing happens. 
\end{itemize}
We say that a customer is \emph{upset} if he chooses a queue whose size exceeds some bound $\varphi$. We are interested in large deviation bounds on the number of upset customers for a given short time interval (say, of $O(D)$ or $\poly\log(D)$ time steps).


Supermarket problems are traditionally considered in the continuous time setting~\cite{M01,VDK96,ELZ86}. But there exists a standard connection between the continuous model and its discrete time counterpart:  conditioned on the number of events is known, the continuous time model behaves 
in the same way as the discrete time counterpart (with parameters appropriately rescaled). 

Most of the existing works~\cite{M01,VDK96,ELZ86} study only the stationary behavior of the processes, such as the expected waiting time and the maximum load among the queues over the time. Here, we are interested in large deviation bounds on a statistics over a \emph{short} time interval; 
the configurations of different cashiers across the time is highly correlated. 

For our purpose, we analyze only the simple special case where the
number of choice $d = 1$; \ie each new customer is put in a random
queue.

We provide a large deviation bound for the number of upset customers
in this setting.\footnote{It is not hard to see that with $D$ cashiers, probability parameter $\alpha$, and ``upset'' threshold $\varphi$, the expected number of upset customers is at most $(\alpha/(1-\alpha))^{\varphi} \cdot T$ for any $T$ steps time interval.}
.
\begin{proposition}\label{prop:supermarket} For the (discrete-time) supermarket problem with $D$ cashier, one choice (i.e., $d=1$), probability parameter $\alpha \in (0,1/2)$, and upset threshold $\varphi \in \N$, for any $T$ steps time interval $[t+1,t+T]$, let $F$ be the number of upset customers in this time interval. We have
\begin{equation}
\Pr{F \geq (1+\delta)(\alpha/(1-\alpha))^{\varphi}T}\leq \left\{\begin{array}{ll}
\exp\left\{-\Omega\left(\frac{\delta^2(\alpha/(1-\alpha))^{\varphi}T}{(1-\alpha)^2}\right) \right\} & \mbox{for } 0 \leq \delta \leq 1\\
\exp\left\{-\Omega\left(\frac{\delta (\alpha/(1-\alpha))^{\varphi}T)}{(1-\alpha)^2}\right) \right\} & \mbox{for }  \delta \geq 1\\
\end{array}
\right.
\end{equation}
\end{proposition}
Note that Proposition~\ref{prop:supermarket} would trivially follow
from the standard Chernoff bound if $T$ is sufficiently large (\i.e.,
$T \gg O(D)$) to
guarantee that we \emph{individually} get concentration on each of the $D$
queue (and then relying on the union bound). What makes Proposition~\ref{prop:supermarket} interesting is that it applies also in a
setting when $T$ is $\poly\log D$.

The proof of Proposition~\ref{prop:supermarket} is found in Section~\ref{sec:step2} and relies on a new variant Chernoff bounds for Markov chains with ``resets,'' which may be of independent interest. 


\begin{Remark} \label{rmk:discrete-to-cont}
One can readily translate the above result to an analogous deviation
bound on the number of upset customers for (not-too-short) time
intervals in the continuous time model.  This follows by noting that
the number of events that happen in a time interval is highly
concentrated (provided that the expected number of events is not too
small), and applying the above proposition after conditioning on the
number of events happen in the time interval (since conditioned on the
number of events, the discrete-time and continous-time processes are identical).
\end{Remark}

\subsection{From ORAM to Supermarkets}\label{sec:step1}
This section shows how we may apply Proposition~\ref{prop:supermarket} to prove Lemma~\ref{lem:queue}. 
Central to our analysis is a simple reduction from the execution of our ORAM algorithm at level $k$ in $\Tr$ to a supermarket process with $D = 2^{k+1}$ cashiers. 
More precisely, we show there exists a coupling between two processes so that each bucket corresponds with two cashiers; 
the load in a bucket is always upper bounded by the total number of customers in the two cashiers it corresponds to. 

To begin, we need the following Lemma.

\begin{Lem}\label{lem:indep}Let $\{a_i\}_{i \geq 1}$ be the sequence of $\proc{Put-Back}$/$\proc{Flush}$ operations defined by our algorithm, 
\ie each $a_i \in \{\proc{Put-Back}, \proc{Flush}\}$ and between any consecutive 
$\proc{Put-Back}$s, the number of $\proc{Flush}$es is a geometric r.v. with parameter $2/3$. 
Then  $\{a_i\}_{i\geq 1}$ is a sequence of i.i.d. random variables so that $\Pr{a_i = \proc{Put-Back}} = \frac 1 3$. 
\end{Lem}

To prove Lemma~\ref{lem:indep}, 
we may view the generation of $\{a_i\}_{i \geq 1}$ as generating a sequence of i.i.d. Bernoulli r.v. $\{b_i\}_{i \geq 1}$ with parameter $\frac 2 3$. We set  $a_i$ be a $\proc{Flush}()$ if and only if $b_i = 1$. One can verify that the $\{a_i\}_{i\geq 1}$ generated in this way is the same as those generated by the algorithm. 

We are now ready to describe our coupling between the original process and the supermarket process. 
At a high-level, a block corresponds to a customer, and $2^{k+1}$ sub-trees in level $k+1$ of $\Tr$ corresponds to $D = 2^{k+1}$ cashiers. More specifically, we couple the configurations at the $k$-th level of $\Tr$ in the ORAM program  with a supermarket process as follows.
\begin{itemize}
\item Initially, all cashiers have $0$ customer.
\item For each $\proc{Put-Back}()$, a corresponding arrival event occurs: if a ball $b$ with position $p = (\gamma||\eta)$ (where $\gamma \in \zo^{k+1}$) is moved to $\Tr$, then a new customer arrives at the $\gamma$-th cashier; otherwise (\eg when the queue is empty), a new customer arrives at a random cashier.
\item For each $\proc{Flush}()$  along the path to leaf $p^* = (\gamma||\eta)$ (where $\gamma \in \zo^{k+1}$), a serving event occurs at the $\gamma$-th cashier.
\item For each $\proc{Fetch}()$, nothing happens in the experiment of the supermarket problem. (Intuitively,  $\proc{Fetch}()$ translates to extra ``deletion'' events of customers in the supermarket problem, but we ignore it in the coupling since it only decreases the number of blocks in buckets in $\Tr$.)
\end{itemize}

\myparab{Correctness of the coupling.} We shall verify the above way of placing and serving customers exactly gives us a supermarket process.  First recall that both $\proc{Put-Back}$ and $\proc{Flush}$ actions are associated with uniformly random leaves. 
Thus, this corresponds to that at each timestep a random cashier will be chosen. Next by Lemma~\ref{lem:indep}, the sequence of $\proc{Put-Back}$ and $\proc{Flush}$ actions in the execution of our ORAM algorithm is a sequence of i.i.d. variables with $\Pr{\proc{Put-Back}} = \frac 1 3$. Therefore, when a queue is chosen at a new timestep, an (independent) biased coin is tossed to decide whether an arrival or a service event will occur. 

\myparab{Dominance.}
Now, we claim that at any timestep, for every $\gamma \in \zo^{k+1}$,
the number of customers at $\gamma$-th cashier is at least the number
of blocks stored  at or above level $k$ in $\Tr$ with position $p =
(\gamma||\cdot)$. This follows by observing that (i) whenever there is
a block with position $p = (\gamma||\cdot)$ moved to $\Tr$ (from
$\proc{Put-Back}()$), a corresponding new customer arrives at the
$\gamma$-th cashier, \ie when the number of blocks increase by one, so
does the number of customers, and (ii) for every $\proc{Flush}()$
along the path to $p^* = (\gamma|| \cdot)$: if there is at least one
block stored at or above level $k$ in $\Tr$ with position $p =
(\gamma||\cdot)$, then one such block will be flushed down below level
$k$ (since we flush the blocks that can be pulled down the
furthest)---that is, when the number of customers decreases by one, so does the number of blocks (if possible). 
This in particular implies that throughout the coupled experiments, for every $\gamma \in \zo^{k}$ the number of blocks in the bucket at node $\gamma$ is always upper bounded by the sum of the number of customers at cashier $\gamma0$ and $\gamma1$. 

We summarize the above in the following lemma.

\begin{Lem}\label{lem:reduction} For every execution of our ORAM algorithm (i.e., any sequence of $\proc{Read}$ and $\proc{Write}$ operations), there is a coupled experiment of the supermarket problem such that throughout the coupled experiments, for every $\gamma \in \zo^{k}$ the number of blocks in the bucket at node $\gamma$ is always upper bounded by the sum of the number of customers at cashier $\gamma0$ and $\gamma1$.
\end{Lem}

\myparab{From Lemma~\ref{lem:reduction} and Proposition~\ref{prop:supermarket} to Lemma~\ref{lem:queue}.}
Note that at any time step $t$, if the queue size is $\leq \frac{1}{2}
\log^{2+\epsilon} n$, then by Proposition~\ref{prop:supermarket} with
$\varphi = \ell/2 = O(\log \log n)$ and
Lemma~\ref{lem:reduction}, except with negligible probability, at time
step $t+ \log^3 n$, there have been at most $\omega(\log n)$ overflows
per level in the tree and thus at most $\frac{1}{2}
\log^{2+\epsilon} n$ in total. Yet during this time ``epoch'', $\log^3 n$
element have been ``popped'' from the queue, so, except with
negligible probability, the queque size cannot exceed $\frac{1}{2}
\log^{2+\epsilon} n$.

It follows by a union bound over $\log^3 n$ length time ``epochs'',
that except with negligible probability, the queue size never exceeds $\log^{2+\epsilon} n$. 

\subsection{Analysis of the Supermarket Problem}\label{sec:step2}
We now prove Proposition~\ref{prop:supermarket}. 
We start with interpreting the dynamics in our process as evolutions
of a Markov chain. 

\myparab{A Markov Chain Interpretation.} In our problem, at each time step $t$, a random cashier is chosen and either an arrival or a serving event happens at that cashier (with probability $\alpha$ and $(1-\alpha)$, respectively), which increases or decreases the queue size by one. Thus, the behavior of each queue is governed by a simple Markov chain  $M$ with state space being the size of the queue (which can also be viewed as a drifted random walk on a one dimensional finite-length lattice). More precisely, each state $i > 0$ of $M$ transits to state $i+1$ and $i-1$ with probability $\alpha$ and $(1-\alpha)$, respectively, and for state $0$, it transits to state $1$ and stay at state $0$ with probability $\alpha$ and $(1-\alpha)$, respectively. In other words, the supermarket process can be rephrased as having $D$ copies of Markov chains $M$, each of which starts from state $0$, and at each time step, one random chain is selected and takes a move.

We shall use Chernoff bounds for Markov chains~\cite{Gillman93,Kahale97,Lezaud04,CLLM12} to derive a large deviation bound on the number of upset customers. Roughly speaking, Chernoff bounds for Markov chains assert that for a (sufficiently long) $T$-steps random walk on an ergodic finite state Markov chain $M$, the number of times that the walk visits a subset $V$ of states is highly concentrated at its expected value $\pi(V) \cdot T$, provided that the chain $M$ has spectral expansion\footnote{For an ergodic reversible Markov chain $M$, the \emph{spectral expansion} $\lambda(M)$ of $M$ is simply the second largest eigenvalue (in absolute value) of the transition matrix of $M$. The quantity $1-\lambda(M)$ is often referred to as the spectral gap of $M$.}
$\lambda(M)$ bounded away from $1$. However, there are a few technical issues, which we address in turn below.

\myparab{Overcounting.} The first issue is that counting the number of
visits to a state set $V \subset S$ does not capture the number of
upset customers exactly---the number of upset customers corresponds to
the \emph{number of transits} from state $i$ to $i+1$ with $i+1 \geq
\varphi$. Unfortunately, we are not aware of Chernoff bounds for
counting the number of transits (or visits to an edge
set). Nevertheless, for our purpose, we can set $V_{\varphi} = \{i: i
\geq \varphi \}$ and the number of visits to $V_{\varphi}$ provides an
\emph{upper bound} on the number of upset customers. 


\myparab{Truncating the chain.} The second (standard) issue is that
the chain $M$ for each queue of a cashier has infinite state space
$\{0\} \cup \N$, whereas Chernoff bounds for Markov chains are only
proven for finite-state Markov chains.  However, since we are only
interested in the supermarket process with finite time steps, we can
simply truncate the chain $M$ at a sufficiently large $K$ (say, $K\gg t+T$) to obtain a chain $M_K$ with finite states $S_K = \{0,1,\dots,K\}$; that is, $M_K$ is identical to $M$, except that for state $K$, it stays at $K$ with probability $\alpha$ and transits to $K-1$ with probability $1-\alpha$. Clearly, as we only consider $t+T$ time steps, the truncated chain $M_K$ behaves identical to $M$. It's also not hard to show that $M_K$ has stationary distribution $\pi_K$ with $\pi_K(i) = (1-\beta)\beta^i/(1-\beta^{K+1})$, and spectral gap $1-\lambda(M_K) \geq \Omega(1/(1-\alpha)^2)$.\footnote{One can see this by lower bounding the conductance of $M_K$ and applying Cheeger's inequality.} 

\myparab{Correlation over a short time frame.} The main challenge,
however, is to establish large deviation bounds for a \emph{short}
time interval $T$ (compared to the number $D$ of chains). For example,
$T = O(D)$ or even $\poly\log(D)$, and in these cases the expected
number of steps each of the $D$ chains take can be a small constant or
even $o(1)$. Therefore, we cannot hope to obtain meaningful
concentration bounds individually for each single chain. Finally, the $D$ chains are not completely independent: only one chain is selected at each time step. This further introduces correlation among the chains. 


We address this issue by relying on a new variant of Chernoff bounds for Markov chains with ``resets,'' which allows us to ``glue'' walks on $D$ separate chains together and yields a concentration bound that is as good as a $T$-step random walk on a single chain. We proceed in the following steps.
\begin{itemize}
\item Recall that we have $D$ copies of truncated chains $M_K$
  starting from state $0$. At each time step, a random chain is
  selected and we takes one step in this chain. We want to upper bound the total number of visits to $V_{\varphi}$ during time steps $[t+1,t+T]$.
\item We first note that, as we are interested in upper bounds, we can
  assume that the chains start at the stationary distribution $\pi_K$
  instead of the $0$ state (i.e., all queues have initial size drawn
  from $\pi_K$ instead of being empty). This follows by noting that
  starting from $\pi_K$ can only increase the queue size
  \emph{throughout} the whole process for \emph{all} of $D$ queues,
  compared to starting from empty queues, and thus the number of
  visits to $V_{\varphi}$ can only increase when starting from $\pi_K$
  in compared to starting from state $0$ (this can be formalized using
  a standard coupling argument).
\item Since walks from the stationary distribution remain at the
  stationary distribution, we can assume w.l.o.g. that the time
  interval is $[1,T]$. Now, as a thought experiment, we can decompose the
  process as follows. We first determine the number of steps each of
  the $D$ chains take during time interval $[1,T]$; let $c_j$ denote
  the number of steps taken in the $j$-th chain. Then we take $c_j$
  steps of random walk from the stationary distribution $\pi_K$ for each copy of the chain $M_K$, and count the total number of visit to $V_{\varphi}$. 
\item Finally, we can view the process as taking a $T$-step random
  walk on $M_K$ with ``resets.'' Namely, we start from the stationary
  distribution $\pi_K$, take $c_1$ steps in $M_K$, "reset" the
  distribution to stationary distribution (by drawing an independent
  sample from $\pi_K$) and take $c_2$ more steps, and so on. At the end, we count the number of visits to $V_{\varphi}$, denoted by $X$, as an upper bound on the number of upset customers. 
\end{itemize}

Intuitively, taking a random walk with resets injects additional
randomness to the walk and thus we should expect at least as good
concentration results. We formalize this intuition as the following
Chernoff bound for Markov chains with "resets"---the proof of which
follows relatively easy from recent Chernoff bounds for Markov chains
\cite{CLLM12} and is found in Appendix~\ref{sec:missing2}---and use it to finish the proof of Proposition~\ref{prop:supermarket}.


\begin{Thm}[Chernoff Bounds for Markov Chains with Resets] \label{thm:Chernoff-reset} Let $M$ be an ergodic finite Markov chain with state space $S$, stationary distribution $\pi$, and spectral expansion $\lambda$. Let $V \subset S$ and $\mu = \pi(V)$. Let $T,D \in \N$ and $1 =  T_0 \leq T_1 \leq \cdots \leq T_D <  T_{D+1} = T+1$. Let $(W_1,\dots,W_T)$ denote a $T$-step random walk on $M$ from stationary with resets at steps $T_1,\dots,T_D$; that is, for every $j \in \{0,\dots,D\}$, $W_{T_j} \leftarrow \pi$ and $W_{T_j+1},\dots,W_{T_{j+1}-1}$ are random walks from $W_{T_j}$. Let $X_i = 1$ iff $W_i \in V$ for every $i \in [T]$ and $X = \sum_{i=1}^T X_i$. We have
$$
\Pr{ X \geq (1+\delta) \mu T}  \leq 
\begin{cases}
\exp\left\{-\Omega(\delta^2  (1-\lambda) \mu T\right)\} & \mbox{ for $0 \leq \delta \leq 1$} \\
 \exp\left\{-\Omega(\delta   (1-\lambda) \mu T \right)\} & \mbox{ for $\delta > 1$}
\end{cases}
$$
\end{Thm}

Now, recall that $1-\lambda(M_K) = \Omega(1/(1-\alpha)^2)$ and  $\pi_K(\varphi) =  \beta^{\varphi}/(1-\beta^{K+1}) = (\alpha/1-\alpha)^{\varphi}/(1-\beta^{K+1})$. Theorem~\ref{thm:Chernoff-reset} says that for every possible $c_1,\dots,c_D$ (corresponding to resetting time $T_j = \sum_{l=1}^j c_j + 1$),
$$
\Pr{ \left.  X \geq \frac{(1+\delta)(\alpha/1-\alpha)^{\varphi}T}{(1-\beta^{K+1})} \right| c_1,\dots,c_D }  \leq 
\begin{cases}
\exp\left\{-\Omega\left(\frac{\delta^2(\alpha/1-\alpha)^{\varphi}T}{(1-\alpha)^2(1-\beta^{K+1})}\right) \right\} & \mbox{for } 0 \leq \delta \leq 1 \\
\exp\left\{-\Omega\left(\frac{\delta (\alpha/1-\alpha)^{\varphi}T)}{(1-\alpha)^2(1-\beta^{K+1})}\right) \right\} & \mbox{for }  \delta \geq 1\\
\end{cases}
$$
Since  $X$ is an upper bound on the number of upset customers, and the above bound holds for every $c_1,\dots,c_D$ and for every $K \geq t+T$, Proposition~\ref{prop:supermarket} follows by taking $K \rightarrow \infty$.  \qed

\vspace{4pt}

\bibliographystyle{plain}
\bibliography{cryptoCornell,CryptoCitations}

\begin{thebibliography}{10}

\bibitem{A10}
Mikl{\'o}s Ajtai.
\newblock Oblivious rams without cryptogrpahic assumptions.
\newblock In {\em STOC}, pages 181--190, 2010.

\bibitem{BMP11}
Dan Boneh, David Mazieres, and Raluca~Ada Popa.
\newblock Remote oblivious storage: Making oblivious ram practical, 20121,
  howpublished = {CSAIL Technical Report: MIT-CSAIL-TR-2011-018}.

\bibitem{CLLM12}
K.~M. Chung, H.~Lam, Z.~Liu, and M.~Mitzenmacher.
\newblock Chernoff-{H}oeffding bounds for {M}arkov chains: Generalized and
  simplified.
\newblock In {\em Proceedings of the 29th International Symposium on
  Theoretical Aspects of Computer Science (STACS)}, 2012.

\bibitem{CP13ORAM}
Kai-Min Chung and Rafael Pass.
\newblock A simple oram.
\newblock Cryptology ePrint Archive, Report 2013/243, 2013.

\bibitem{DMN11}
Ivan Damg{\aa}rd, Sigurd Meldgaard, and Jesper~Buus Nielsen.
\newblock Perfectly secure oblivious ram without random oracles.
\newblock In {\em TCC}, pages 144--163, 2011.

\bibitem{ELZ86}
Derek~L. Eager, Edward~D. Lazowska, and John Zahorjan.
\newblock Adaptive load sharing in homogeneous distributed systems.
\newblock {\em IEEE Trans. Software Eng.}, 12(5):662--675, 1986.

\bibitem{GGHJRW13}
Craig Gentry, Kenny~A. Goldman, Shai Halevi, Charanjit~S. Jutla, Mariana
  Raykova, and Daniel Wichs.
\newblock Optimizing oram and using it efficiently for secure computation.
\newblock In {\em Privacy Enhancing Technologies}, pages 1--18, 2013.

\bibitem{Gillman93}
D.~Gillman.
\newblock A chernoff bound for random walks on expander graphs.
\newblock {\em SIAM Journal on Computing}, 27(4), 1997.

\bibitem{G87}
Oded Goldreich.
\newblock Towards a theory of software protection and simulation by oblivious
  rams.
\newblock In {\em STOC}, pages 182--194, 1987.

\bibitem{GO96}
Oded Goldreich and Rafail Ostrovsky.
\newblock Software protection and simulation on oblivious rams.
\newblock {\em J. ACM}, 43(3):431--473, 1996.

\bibitem{GM11}
Michael~T. Goodrich and Michael Mitzenmacher.
\newblock Privacy-preserving access of outsourced data via oblivious ram
  simulation.
\newblock In {\em ICALP (2)}, pages 576--587, 2011.

\bibitem{GMOT12}
Michael~T. Goodrich, Michael Mitzenmacher, Olga Ohrimenko, and Roberto
  Tamassia.
\newblock Privacy-preserving group data access via stateless oblivious ram
  simulation.
\newblock In {\em SODA}, pages 157--167, 2012.

\bibitem{Kahale97}
N.~Kahale.
\newblock Large deviation bounds for markov chains.
\newblock {\em Combinatorics, Probability, and Computing}, 6(4), 1997.

\bibitem{KLO12}
Eyal Kushilevitz, Steve Lu, and Rafail Ostrovsky.
\newblock On the (in)security of hash-based oblivious ram and a new balancing
  scheme.
\newblock In {\em SODA}, pages 143--156, 2012.

\bibitem{Lezaud04}
P.~Lezaud.
\newblock Chernoff-type bound for finite markov chains.
\newblock {\em Annals of Applied Probability}, 8(3):849--867, 1998.

\bibitem{M01}
Michael Mitzenmacher.
\newblock The power of two choices in randomized load balancing.
\newblock {\em IEEE Trans. Parallel Distrib. Syst.}, 12(10):1094--1104, 2001.

\bibitem{MPS02}
Michael Mitzenmacher, Balaji Prabhakar, and Devavrat Shah.
\newblock Load balancing with memory.
\newblock In {\em FOCS}, pages 799--808, 2002.

\bibitem{MV99}
Michael Mitzenmacher and Berhold Vocking.
\newblock The asymptotics of selecting the shortest of two, improved.
\newblock In {\em PROCEEDINGS OF THE ANNUAL ALLERTON CONFERENCE ON
  COMMUNICATION CONTROL AND COMPUTING}, volume~37, pages 326--327, 1999.

\bibitem{OS97}
Rafail Ostrovsky and Victor Shoup.
\newblock Private information storage (extended abstract).
\newblock In {\em STOC}, pages 294--303, 1997.

\bibitem{PR10}
Benny Pinkas and Tzachy Reinman.
\newblock Oblivious ram revisited.
\newblock In {\em CRYPTO}, pages 502--519, 2010.

\bibitem{DB02}
Devavrat Shah and Balaji Prabhakar.
\newblock The use of memory in randomized load balancing.
\newblock In {\em Information Theory, 2002. Proceedings. 2002 IEEE
  International Symposium on}, page 125. IEEE, 2002.

\bibitem{SCSL11}
Elaine Shi, T.-H.~Hubert Chan, Emil Stefanov, and Mingfei Li.
\newblock Oblivious ram with o((logn)3) worst-case cost.
\newblock In {\em ASIACRYPT}, pages 197--214, 2011.

\bibitem{SS12}
Emil Stefanov and Elaine Shi.
\newblock Path o-ram: An extremely simple oblivious ram protocol.
\newblock {\em CoRR}, abs/1202.5150v1, 2012.

\bibitem{SSS12}
Emil Stefanov, Elaine Shi, and Dawn Song.
\newblock Towards practical oblivious ram.
\newblock In {\em NDSS}, 2012.

\bibitem{SvDSFRYD13}
Emil Stefanov, Marten van Dijk, Elaine Shi, Christopher Fletcher, Ling Ren,
  Xiangyao Yu, and Srinivas Devadas.
\newblock Path o-ram: An extremely simple oblivious ram protocol.
\newblock {\em CoRR}, abs/1202.5150v2, 2013.

\bibitem{VDK96}
Nikita~Dmitrievna Vvedenskaya, Roland~L'vovich Dobrushin, and
  Fridrikh~Izrailevich Karpelevich.
\newblock Queueing system with selection of the shortest of two queues: An
  asymptotic approach.
\newblock {\em Problemy Peredachi Informatsii}, 32(1):20--34, 1996.

\bibitem{WS08}
Peter Williams and Radu Sion.
\newblock Usable pir.
\newblock In {\em NDSS}, 2008.

\bibitem{WSC08}
Peter Williams, Radu Sion, and Bogdan Carbunar.
\newblock Building castles out of mud: practical access pattern privacy and
  correctness on untrusted storage.
\newblock In {\em ACM Conference on Computer and Communications Security},
  pages 139--148, 2008.

\end{thebibliography}

\appendix

\section{Implementation details.}\label{sec:impl}
This section discusses a number of implementation details in our algorithm. 

\myparab{The queue at the cache.} We now describe how we may use a hash table and a standard queue (that 
could be encapsulated in commodity chips) to implement our queue. Here, we only assume the hash table uses universal hash function and it resolves collisions by using a linked-list. To implement the $\proc{Insert}(\mathrm{Block:} b)$ procedure, we simply insert $b$ to both the hash table and the queue. The key we use is $b$'s value at the position map. Doing so we may make sure the maximum load of the hash table is $O(\log n)$ whp~\cite{MV08}. To implement $\proc{Find}(\mathrm{int:} i, \mathrm{word:} p)$, we find the block $b$ from the hash table. If it exists, return the block and delete it. Notice that we \emph{do not} delete $b$ at the queue. So this introduces inconsistencies between the hash table and the queue. 

We now describe how we implement $\proc{PopFront}()$. Here, we need to be careful with the inconsistencies. We first pop a block from the queue. Then we need to check whether the block is in hash table. If not, that means the block was already deleted earlier. In this case, $\proc{PopFront}()$ will not return anything (because we need a hard bound on the running time). One can see that $\proc{Insert}()$ takes $O(1)$ time and the other two operations take $\omega(\log n)$ time whp. 

\myparab{The $\proc{Flush()}$ and $\proc{Overflow}$() procedures.} Figure~\ref{fig:flush} and Figure~\ref{fig:overflow} also give pseudocode for the $\proc{Flush()}$ and $\proc{Overflow}$() procedures.

\section{Missing Proofs}
\subsection{Proof of Lemma \ref{lem:leaf}}
\begin{proof}
We turn to showing that the probability of overflow in any of the leaf
nodes is small. 
Consider any leaf node $\gamma$ and some time $t$. For there to be
an overflow in $\gamma$ at time $t$, there must be $\ell'+1$ out of $n/\alpha$
elements in the position map that map to $\gamma$. Recall that all
positions in the position map are uniformly and independently
selected; thus, the expected number of elements mapping to $\gamma$ is
$\mu=\log n \log \log n$ and by a standard multiplicative version of Chernoff bound, the
probability that $\ell'+1$ elements are mapped to $\gamma$ is upper bounded
by $2^{-\ell'}$ when $\ell' \geq 6 \mu$ (see Theorem 4.4 in~\cite{mitzenmacher2005probability}). 
By a union bound, we have that the probability of \emph{any} node ever
overflowing is bounded by 
$2^{-(\ell')} \cdot (n/\alpha)\cdot T$

To analyze the full-fledged construction, we simply apply the union
bound to the failure
probabilities of the $\log_{\alpha} n$ different ORAM trees  (due to the recursive
calls). 
The final upper bound on the overflow probability is thus
$2^{-(\ell')} \cdot (n/\alpha)\cdot T \cdot \log_{\alpha} n$, which is
negligible as long as $\ell' = c\log n \log \log n$ for a suitably large constant $c$.
\end{proof}

\subsection{Proof of Theorem \ref{thm:Chernoff-reset}}\label{sec:missing2}

We here prove Theorem~\ref{thm:Chernoff-reset}. The
high level idea is simple---we simulate the resets by taking a
sufficiently long ``dummy'' walk, where we ``turn off'' the counter on
the number of visits to the state set $V$. However, formalizing this
idea requires a more general version of Chernoff bounds that handles
``time-dependent weight functions,'' which allows us to turn on/off
the counter. Additionally, as we need to add long dummy walks, a
multiplicative version (as opposed to an additive version) Chernoff
bound is needed to derive meaningful bounds. We here rely on a recent
generalized version of Chernoff bounds for Markov chains due to Chung,
Lam, Liu and Mitzenmacher~\cite{CLLM12}.

\begin{Thm}[\cite{CLLM12}] \label{thm:mixdeviation} Let $M$ be an ergodic finite Markov chain with state space $S$, stationary distribution $\pi$, and spectral expansion $\lambda$. Let $\mathcal W = (W_1,\dots, W_T)$ denote a $T$-step random walk on $M$ starting from stationary distribution $\pi$, that is,  $W_1 \leftarrow \pi$. For every $i \in [T]$, let $f_i: S \rightarrow [0,1]$ be a weight function at step $i$ with expected weight $\E_{v\leftarrow \pi}[f_i(v)] = \mu_i$. Let $\mu = \sum_i \mu_i$. Define the total weight of the walk $(W_1,\dots, W_t)$ by $X \triangleq \sum_{i=1}^t f_i(W_i)$. Then
$$
\Pr{ X \geq (1+\delta) \mu}  \leq 
\begin{cases}
\exp\left\{ -\Omega(\delta^2  (1-\lambda) \mu )\right\} & \mbox{ for $0 \leq \delta \leq 1$} \\
 \exp\left\{ -\Omega(\delta   (1-\lambda) \mu ) \right\} & \mbox{ for $\delta > 1$}
\end{cases}
$$
\end{Thm}

We now proceed to prove Theorem~\ref{thm:Chernoff-reset}. 

\begin{proof}[Proof of Theorem~\ref{thm:Chernoff-reset}] We use Theorem~\ref{thm:mixdeviation} to prove the theorem. Let $f: S \rightarrow [0,1]$ be an indicator function on $V \subset S$ (i.e., $f(s) = 1$ iff $s\in V$) .The key component from Theorem~\ref{thm:mixdeviation} we need to leverage here is that the functions $f_i$ can change over the time. Here, we shall design a very long walk $\mathcal V$ on $M$ so that the marginal distribution of a specific collections of ``subwalks'' from $\mathcal V$ will be statistically
close to $\mathcal W$. Furthermore, we design $\{f_i\}_{i \geq 0}$ in such a way that those ``unused'' 
subwalks will have little impact to the statistics we are interested in. 
In this way, we can translate a deviation bound on $\mathcal V$ to a deviation  bound on $\mathcal W$. Specifically,
let $T(\epsilon)$ be the mixing time for $M$ (\ie the number of steps needed for a walk to be $\epsilon$-close to the stationary distribution in statistical distance). Here, we let $\epsilon \triangleq \exp(-DT)$ ($\epsilon$ is chosen in an arbitrary manner so long as it is sufficiently small).
 Given $1 =  T_0 \leq T_1 \leq \cdots \leq T_D <  T_{D+1} = T+1$, 
 we define $\mathcal V$ and $f_i$ as follows: $\mathcal V$ will start from $\pi$ and take $T_1 -2$ steps of walk. In the mean time, we shall set $f_i = f$ for all $i < T_1$. Then we ``turn off'' the function $f_i$ while letting $\mathcal V$ keep walking for $T(\epsilon)$ more steps, \ie we let 
$f_i = 0$ for all $T_1 \leq i \leq T_1 + T(\epsilon) - 1$. Intuitively, this means we let $\mathcal V$ take a long walk until it becomes close to $\pi$ again. During this time, $f_i$ is turned off so that we do not keep track of any statistics. 
After that, we ``turn on'' the function $f_i$ again for the next $T_2 - T_1$ steps (\ie $f_i = f$ for all $T_1 + T(\epsilon)  \leq  i \leq T_2 + T(\epsilon) - 1$, followed by turning $f_i$ off for another $T(\epsilon)$ steps. We continue this ``on-and-off'' process until we walk through all $T_j$'s. 

Let $\mathcal V'$ be the subwalks of $\mathcal V$ with non-zero $f_i$. One can see that the statistical distance between $\mathcal V'$ and $\mathcal W$ is $\mathrm{poly}(D, T)\exp(-DT) \leq \exp(-T+o(T))$. Thus, for any $\theta$ we have
\begin{equation}\label{eqn:bound}
\Pr{\sum_{w \in \mathcal W}f(w) \geq \theta} \leq \Pr{\sum_{v' \in \mathcal V'}f(v') \geq \theta} + \exp(-T+o(T))  = \Pr{\sum_{v \in \mathcal V}f(v) \geq \theta} + \exp(-T+o(T)).
\end{equation}
By letting $\theta = (1+\delta)\mu T$ and using Theorem~\ref{thm:mixdeviation} to the right hand side of (\ref{eqn:bound}), we finish our proof. 
\end{proof}

\begin{figure}
\begin{codebox}
\Procname{$\proc{Flush}(\Tr)$}
\li Let $p^*$ be a uniformly random leaf. 
\li Denote $\tpath(\rt, p^*)$ as $p_0p_1...p_{d }$ (with $p_0 = \rt$ and $p_{d } = p^*$). 
\li $\mathrm{block}\gets \mathrm{null}.$
\li \For $i \gets 0$ \To $d - 1$
\li \Do 
\li Read the buckets $\vec b$ at node $p_i$  to the client side.  
\li \If $\mathrm{block}\neq \mathrm{null}$
\li \Then Insert $\mathrm{block}$ to $\vec b$ by replacing it with a dummy block in $\vec b$.  
\li Find a block $b[j] \in \vec b$ such that    
$p_{i + 1} \in \tpath(r, P[b[j]])$. 
\li \Comment If there are more than one such blocks, 
\li \Comment \quad find the one that can travel furtherest.  
\li \If Can find such $b[j]$
\li \Then
$\mathrm{block}\gets b[j]$ 
\li Replace $b[j]$ by a dummy block 
\li
\Else $\mathrm{block} \gets \mathrm{null}$. 
\End
\li Let $S_L, S_R \subset \vec b$ be set of balls belong to left and right sub-trees.
\li \If $|S_L| \geq \ell/2$ 
\li \Then 
select any  $b \in S_L$ and replace it by a dummy ball. 
\li $\proc{Overflow}(b)$
\End
\li \If $|S_R| \geq \ell/2$ 
\li \Then 
select any  $b \in S_R$ and replace it by a dummy ball. 
\li $\proc{Overflow}(b)$
\End
\li Write back $\vec b$ to $p_i$.
\End
\End
\li Read the blocks $\vec b$ at $p_d$.  
\li \If $\mathrm{block}\neq \mathrm{null}$
\li \Then Insert $\mathrm{block}$ to $\vec b$ by replacing it with a dummy block in $\vec b$.  \End
\li \If $\vec b$ is full 
\li \Then abort the program. \End
\li Write back $\vec b$ to $p_d$.
\end{codebox}
\caption{Pseudocode for the $\proc{Flush}(\cdot)$ action}
\label{fig:flush}
\end{figure}

\begin{figure}
\begin{codebox}
\Procname{$\proc{Overflow}(\mathrm{block:} b)$}
\li Update $P[i]$ to a uniformly random value from $[L]$, where $i$ is the index of $b$. 
\li Update the position $p = P[i]$ in $b$. 
\li Insert $b$ to the queue. 
\end{codebox}
\caption{Pseudocode for the $\proc{Overflow}$ procedure}
\label{fig:overflow}
\end{figure}

\end{document}